\documentclass[aps,prx,twocolumn,10pt,superscriptaddress,showkeys,floatfix]{revtex4-1}

\usepackage{lmodern}
\usepackage[utf8]{inputenc}
\usepackage{graphicx}
\usepackage{longtable}
\usepackage{dcolumn}
\usepackage{bm}
\usepackage{xcolor}
\usepackage{amssymb}
\usepackage{hyperref}
\usepackage{tabularx}

\usepackage{newtxtext} 
\usepackage{amsmath}
\usepackage[bigdelims]{newtxmath}
\usepackage[T1]{fontenc}
\usepackage{textcomp}

\newcommand{\YBCO}{YBa$_2$Cu$_3$O$_{y}$}

\newcommand{\YBCOopt}{YBa$_2$Cu$_3$O$_{6.9}$}
\newcommand{\YCaBCO}{Y$_{1-x}$Ca$_x$Ba$_2$Cu$_3$O$_{y}$}
\newcommand{\YSCO}{YSr$_2$Cu$_3$O$_y$}
\newcommand{\CuMo}{Cu$_{1-x}$Mo$_x$Sr$_2$YCu$_2$O$_y$}
\newcommand{\BCO}{Ba$_2$CuO$_{3+y}$}
\newcommand{\SCO}{Sr$_2$CuO$_{3+y}$}
\newcommand{\LCO}{La$_2$CuO$_4$}
\newcommand{\LSCO}{La$_{2-x}$Sr$_x$CuO$_4$}
\newcommand{\LCCO}{La$_{2-x}$Ca$_x$CuO$_4$}
\newcommand{\TBCO}{Tl$_2$Ba$_2$CuO$_{6+y}$}

\begin{document}

\title{Superconductivity in strongly overdoped cuprates: beyond the single-band model}

\author{Ruichao Chen}
\affiliation{IMPMC, Sorbonne Universit\'e, CNRS, MNHN, 4, place Jussieu, 75005 Paris, France}
\author{Linda Sederholm}
\affiliation{Aalto University, Department of Chemistry and Materials Science, Espoo 00076, Finland}
\author{Yannick Klein}
\author{Ludovic Delbes}
\author{Beno\^it Baptiste}
\author{Paraskevas Parisiades}
\affiliation{IMPMC, Sorbonne Universit\'e, CNRS, MNHN, 4, place Jussieu, 75005 Paris, France}
\author{Emmanuel Maisonhaute}
\affiliation{Sorbonne Universit\'e, CNRS, Institut Parisien de Chimie Moléculaire (IPCM), 4, place Jussieu, 75005 Paris, France}
\author{Davide Delmonte}
\author{Edmondo Gilioli}
\affiliation{IMEM-CNR, Parco Area delle Scienze 37/A, 43124, Parma, Italy}
\author{Maarit Karppinen}
\affiliation{Aalto University, Department of Chemistry and Materials Science, Espoo 00076, Finland}
\author{Ronald I. Smith}
\author{David A. Keen}
\affiliation{ISIS Neutron and Muon Source, Rutherford Appleton Laboratory, Didcot, OX11 0QX, UK}
\author{Yann Le Godec}
\author{Andrea Gauzzi}
\email[]{andrea.gauzzi@sorbonne-universite.fr}
\affiliation{IMPMC, Sorbonne Universit\'e, CNRS, MNHN, 4, place Jussieu, 75005 Paris, France}

\date{\today}

\begin{abstract}

In order to explain the observation of an extended superconducting region in several overdoped cuprates, which contrasts the dome scenario, by means of neutron and synchrotron x-ray powder diffraction we study the crystal structure of \YBCO, where strong oxygen overdoping up to $y = 7.4$ is achieved under high-pressure. A bond valence sum analysis indicates that 1/5 of the extra holes created by the excess oxygen are transferred to the CuO$_2$ planes, thus increasing the hole density up to $p=0.27$ hole/Cu, where superconductivity is expected to vanish according to the dome scenario. Instead, our data confirm a previous observation [Okai, Ono and Mitsuhashi, Physica C: Superconductivity {\bf 366}, 164 (2002)] that the superconducting critical temperature, $T_c$, remains constant with $y$. Our data analysis accounts for this discrepancy in terms of the much shorter bond between the apical oxygen and the planar Cu ion, which suggests that the extra holes occupy the $a_1$-symmetry states formed by $d_{3z^2-r^2}$ orbitals, instead of the usual $b_1$-symmetry Zhang-Rice singlet states formed by $d_{x^2-y^2}$ orbitals. Suitable spectroscopic measurements on single crystals may support such a two-band scenario, which would require a totally different theoretical approach to explain superconductivity in cuprates.

\end{abstract}

\maketitle

\section{Introduction}
Owing to a common crystal structure made of CuO$_2$ planes, it is widely assumed that all superconducting cuprates follow the same dome-shaped dependence of the critical temperature, $T_c$, on hole density, $p$, in the planes \cite{zha93,tal95}. For decades, this dependence has been used as phenomenological phase diagram to unveil the still elusive pairing mechanism of cuprates. In fact, only the underdoped side of the dome, where $T_c$ increases with $p$ up to an optimal doping level, $p^{\ast} \approx 0.17$ hole/Cu, has been firmly established. Higher levels require high concentrations of chemical substitutions or oxygen. As a result, the overdoped side at  $p > p^{\ast}$, where $T_c$ decreases down to zero at $p \approx 0.27$ hole/Cu, has been clearly established only in \LSCO\ and \TBCO\ \cite{pre91,tal95}, while only partial overdoping has been achieved in other systems like \YCaBCO\ \cite{awa94,bot96}.

The general validity of the dome picture for all cuprates has been questioned by a growing body of data showing that several systems, including high-pressure oxygenated \YBCO\ \cite{oka02} and \CuMo\ \cite{chm10,gau16}, \SCO\ \cite{hir93,yan07}, \BCO\ \cite{li19} and \LCCO\ \cite{kim21}, display an extended superconducting region at doping levels as high as $p \sim 0.5$ hole/Cu, well beyond the dome, where a conventional Fermi liquid is rather expected. Notably, two of these systems, \YBCO\ and \LCO, exhibit a dome behavior when overdoping is achieved by means of Ca-substitution, instead of oxygen, or Sr-, instead of Ca-substitution, respectively.  

\begin{figure}[htbp]
    \centering
    \includegraphics[width=\columnwidth]{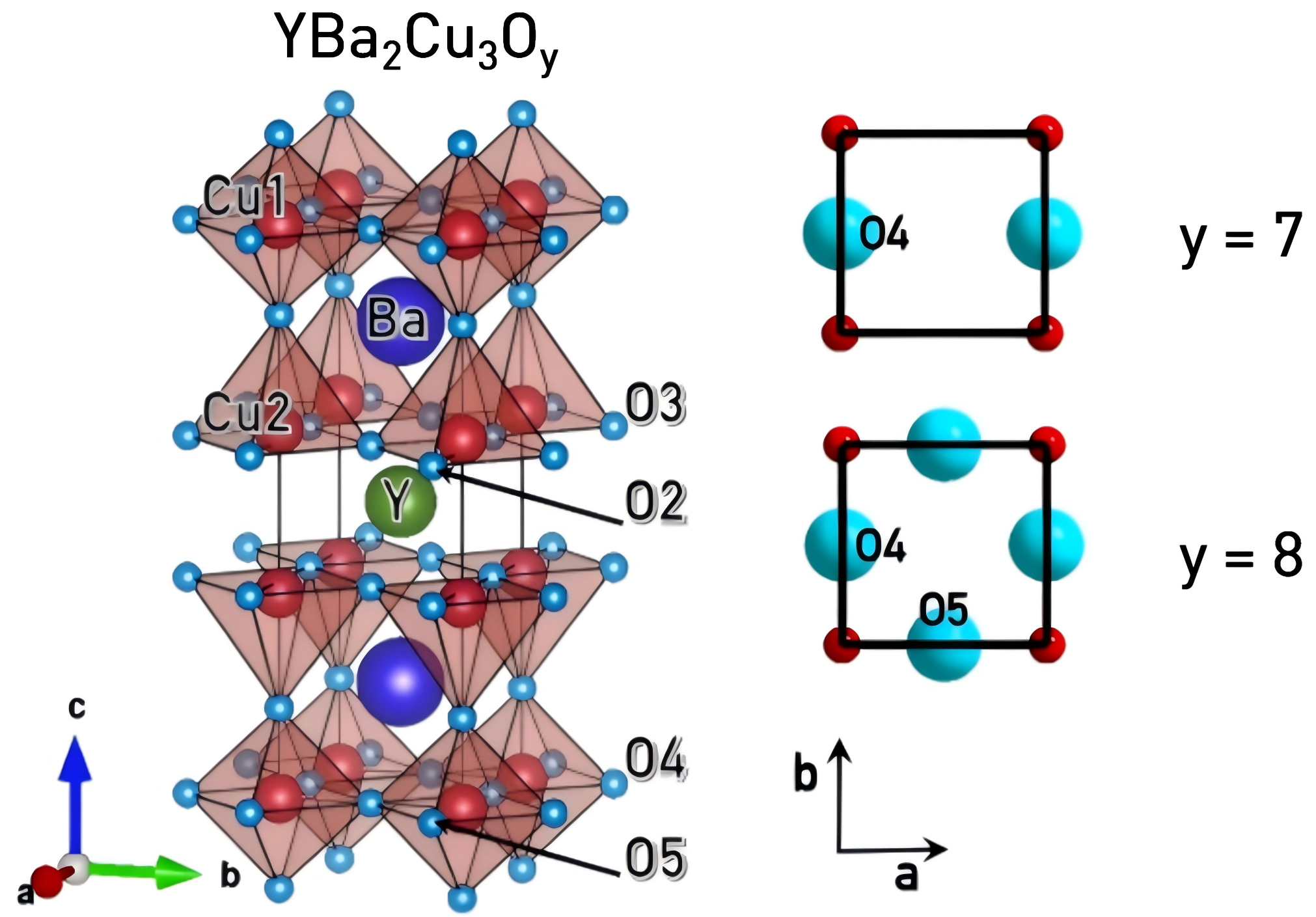} 
    \caption{Left: crystal structure of \YBCO. Site labels follow Ref. \cite{cap87}. Right: occupancy of the O4 and O5 sites in the limiting cases $y=7$, i.e. slightly above optimal doping $y^{\ast}=6.9$, and $y=8$. In the present work we achieved overdoping levels up to $y=7.4$.} 
    \label{fig:structure}
\end{figure}

One possible explanation of this contrasting behavior is that the same level of nominal overdoping may affect differently the hole states in the CuO$_2$ planes, depending on the specific type of dopant employed. For example, \YBCO\ (YBCO) and \CuMo\ contain an additional Cu site in the CuO chains that play the role of charge reservoir layer (see Figure \ref{fig:structure}), so a fraction of holes created by the excess oxygen may remain localized in this layer. To address this point, here we investigate the changes of crystal structure caused by oxygen overdoping in YBCO for $y$ up to 7.4, corresponding to an excess oxygen $\Delta y = 0.5$ with respect to optimal doping, $y^{\ast}=6.9$. We find that these changes are radically different from those caused by Ca overdoping, thus giving a hint to explain the extended superconducting region observed.

\section{Experimental methods}

\subsection{High-pressure oxygenation (HPO)}
\label{HPO}

We achieved high oxygen concentrations by means of a high-pressure oxygenation (HPO) reaction between optimally doped \YBCOopt\ precursor powders and KClO$_3$ or AgO powders used as oxidizing agent, as previously reported \cite{oka90,oka96,gil00}. We prepared the precursor by a standard solid-state method \cite{ans91,and97} which ensures high purities and sharp superconducting transitions with onset $T_c = 93$ K, as indicated by X-ray diffraction (XRD) and magnetization measurements.

The HPO samples are prepared in a Walker-type multi-anvil press of 1000 ton capacity, where the pressure cell is filled with a mixture of YBCO precursor (hereafter, sample P) and KClO$_3$ (or AgO) powders in a proportion corresponding to a nominal composition up to $y = 8.5$, larger than the maximum, $y=8$, that can be hosted by the crystal structure. We obtained single-phase samples using optimized conditions of 4-6 GPa and 475-550 $^{\circ}$C for 3.5 hours. Slightly different conditions led to lower oxygen concentrations and limited ($< 5\%$) impurity concentrations, mostly BaO$_2$. The present study is focused on eight samples, labeled N1-N4 and S1-S4, prepared under the conditions reported in Appendix \ref{app:synthesis}, Tables \ref{tab:HPO_ISIS} and \ref{tab:HPO_SOLEIL}, that we measured by neutron and synchrotron XRD, respectively. 

In order to disentangle the effects of the oxidizing agent from those of high-pressure (HP), we prepared a reference sample (R) by treating the precursor P under the same HP conditions but without oxidizing agent. We therefore expect that the oxygen concentration in sample R is unchanged ($y^{\ast}=6.9$), with respect to the precursor sample, P.   

\subsection{Structure and magnetization characterization}
All HPO samples were characterized by room temperature XRD in the Bragg‐Brentano geometry using a commercial Panalytical X'pert Pro powder diffractometer equipped with a Co K$_{\alpha}$ source. Samples N1-N4 and S1-S4 were selected for a high-resolution structural study by time-of-flight neutron diffraction at the Polaris instrument \cite{smi19} of the ISIS neutron source or by monochromatic ($\lambda$ = 0.51363 \AA) synchrotron XRD at the CRISTAL beamline of the SOLEIL synchrotron.

As a complementary method to estimate $y$, we employed Raman spectroscopy. We used a Horiba HR NanoEvo spectrometer equipped with an Olympus 80x/0.75na long working distance objective to match the high rugosity of the surface and still focus on the surface of individual grains. The laser power was kept below 0.5 mW using neutral density filters to avoid laser heating. The spectral resolution was set to 2 $\mu\mathrm{m}^{-1}$ corresponding to a grating of 1200 grooves/mm.


The superconducting properties were studied by means of DC magnetization measurements in a vibrating sample magnetometer (VSM) of a Physical Property Measurement System (PPMS) or in a VSM-Superconducting Quantum Interference Device (SQUID) apparatus by Quantum Design. Both zero-field-cooling and field-cooling (ZFC, FC) curves were measured in the 2-300 K range in a field of 20 Oe.

\section{Rietveld structure refinements}

\subsection{Neutron diffraction results}

The XRD data confirm the high purity of the N1-N4 samples. Samples N1 and N4 display no impurities within the instrumental resolution. In samples N2 and N3, we detected weak impurity peaks attributed to metallic Ag or BaO$_2$, respectively, corresponding to a molar concentration of 2\% or less. Neutron diffraction data were preferred for crystal structure refinement due to their better resolution and sensitivity to lighter oxygen atoms in the presence of heavier metals. Diffraction patterns collected in the Polaris backscattering detector bank were fitted by the Rietveld method as implemented in the GSAS software \cite{bri13}. For all samples, we successfully used the orthorhombic $Pmmm$ structure that describes \YBCO\ for $y > 6.4$. Following Capponi \textit{et al.} \cite{cap87}, we label Cu1 and Cu2 the chain and planar Cu sites, respectively, O1 the apical oxygen site, O2 and O3 the planar oxygen sites along the $a$- and $b$-axis, respectively, and O4 and O5 the oxygen chain sites along the $b$- and $a$-axis, respectively (see Figure \ref{fig:structure}). In the refinements, we assumed full occupancy of the oxygen O4 site in the CuO chain and varied the occupancy of the initially empty oxygen O5 site. This constraint enhances the stability of the numerical convergence and takes into account the fact that the $b$-axis parameter does not change appreciably with $y$ (see Figure \ref{fig:abc_vs_y}), indicating that the O4 site remains fully occupied, while the $a$-axis parameter increases, as expected assuming the extra oxygen occupies the O5 site. To improve the stability of the numerical convergence, we restricted ourselves to refine the isotropic thermal parameters, $B_{\rm iso}$.

\begin{figure}[htbp]
    \centering
    \includegraphics[width=\columnwidth]{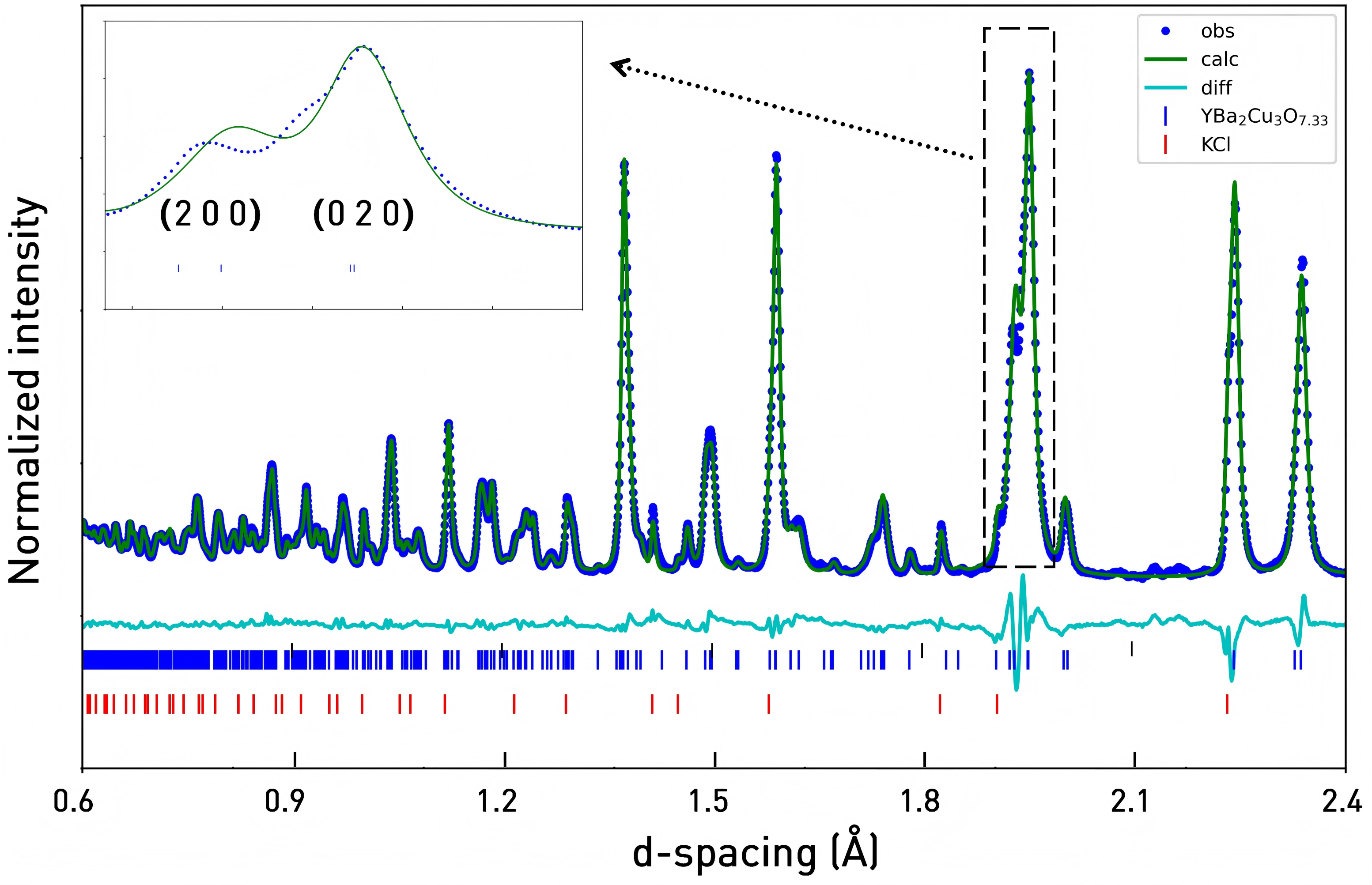} 
    \caption{Observed and calculated  diffraction profiles of sample N1 ($y=0.33$) and their difference. See the refined structure in Table \ref{tab:structure_ISIS}. The data were collected by Bank 5 of the POLARIS diffractometer (see text). Blue and red ticks mark the Bragg peaks of \YBCO\ and KCl, respectively. Inset: detail of the (200) and (020) peaks.} 
    \label{fig:neutron}
\end{figure}

In Table \ref{tab:structure_ISIS} of Appendix \ref{app:refined}, we report the refined structure of the N1-N4 samples. For comparison, we also give the structure of optimally doped YBCO earlier reported by Jorgensen et al. \cite{jor90} and consistent with other reports \cite{cap87,yan87,jor90,sch94}. Note the high quality of the refinements obtained for all N1-N4 samples. This is seen also in Figure \ref{fig:neutron}, where we plot the observed and calculated profiles of the representative sample N1. We note a difference between observed and calculated intensities only in the (200) and (020) peaks, attributed to a deviation of the local structure from the average one. Indeed,  different local configurations of the excess oxygen in the O5 sites are possible. 

\subsection{X-ray synchrotron diffraction results}
We carried out a similar structure refinement for the S1-S4 samples using the XRD data. These samples too are single-phase or contain impurity concentrations less than 5\% (see Table \ref{tab:HPO_SOLEIL}). The only difference in the refinement procedure used for these samples regards the occupancy of the O5 site. Given the lower sensitivity of x-rays to oxygen, refining the site occupancy factor of the O5 site causes instability of the numerical procedure. Hence, we fixed this factor consistent with the oxygen concentration $y$ estimated from the lattice parameters or from the frequency of the A$_{1g}$ Raman mode, as illustrated in the next sections.


\section{Determination of the excess oxygen}

\subsection{Rietveld structure refinement}
A straightforward method to determine $y$ consists in the Rietveld refinement of the occupancy factor of the oxygen O5 site. As seen in Table \ref{tab:structure_ISIS}, we applied this method to the N1-N4 samples and obtained $y=$ 0.33, 0.10, 0.42 and 0.34. Given the quality of the refinements reported in the preceding section, we estimate an uncertainty $\delta y = \pm 0.05$. For the N4 sample, we performed a complementary TGA analysis that confirmed the value $y=0.34$, so we are confident that the Rietveld refinement is a reliable method to determine $y$.

The comparable oxygen concentrations, $y$, in samples N1, N3 and N4 indicate that separating the oxidizing agent from the precursor YBCO powders in the high-pressure cell does not affect $y$. We conclude that, under the HPO conditions used, the oxygen diffusion length is longer than sample dimension. We attribute the larger $y$ of sample N3 to a higher synthesis temperature. The smaller $y$ of sample N2 suggests that AgO is less effective than KClO$_3$ as oxidizing agent. 

\subsection{Dependence of lattice parameters on $y$}

\begin{figure}[htbp]
    \centering
    \includegraphics[width=\columnwidth]{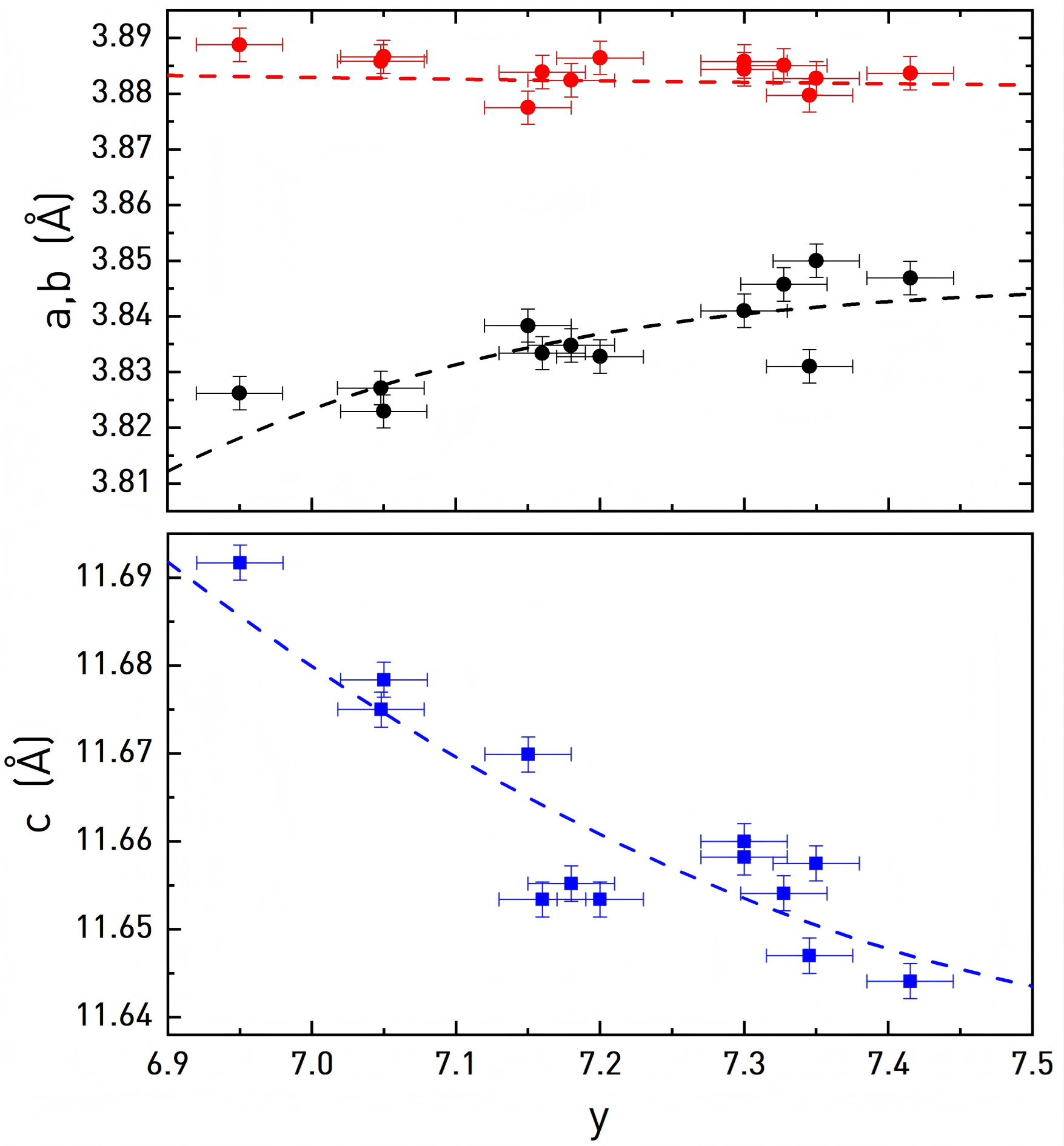} 
    \caption{Dependence of the $a$-, $b$- and $c$-lattice parameters of the \YBCO\ HPO samples studied in this work on oxygen concentration, $y$. Broken lines interpolate the data by Okai \cite{oka02}.
    } 
    \label{fig:abc_vs_y}
\end{figure}

A second method to determine $y$ relies on the experimental dependence of the lattice parameters on $y$ previously established by Okai et al. \cite{oka99,oka02} who determined $y$ by TGA in a series of \YBCO\ samples prepared using similar HPO conditions as ours. We have then interpolated Okai's data to obtain a set of calibration curves, $a$, $b$, and $c$ vs. $y$, and thus to estimate $y$ in our HPO samples. The result is reported in Figure \ref{fig:abc_vs_y}. To verify the reliability of the method, we plotted the $a$, $b$, and $c$ vs. $y$ data obtained independently for samples N1-N4 using the previous Rietveld refinement method and found full agreement with Okai's data. One notes the following:

\noindent
1. The $b$-axis parameter remains nearly constant with $y$, suggesting that the O4 site remains fully occupied, i.e. the CuO chains along the $b$-axis are not affected by oxygen overdoping.

\noindent
2. Consistent with the above observation, the increase of $a$-axis parameter indicates that the initially empty O5 site is progressively occupied with increasing $y$.

\noindent
3. The $c$-axis parameter continues to decrease with $y$, thus extrapolating the trend in the underdoped region \cite{cav90,jor90}. This is expected, for the Cu1-O1 and Cu2-O1 bond lengths decrease with increasing number of oxygen ions bound to the Cu1 ion.

\subsection{Dependence of the A$_{1g}$ Raman shift on $y$}
\begin{figure}[htbp]
    \centering
    \includegraphics[width=0.9\columnwidth]{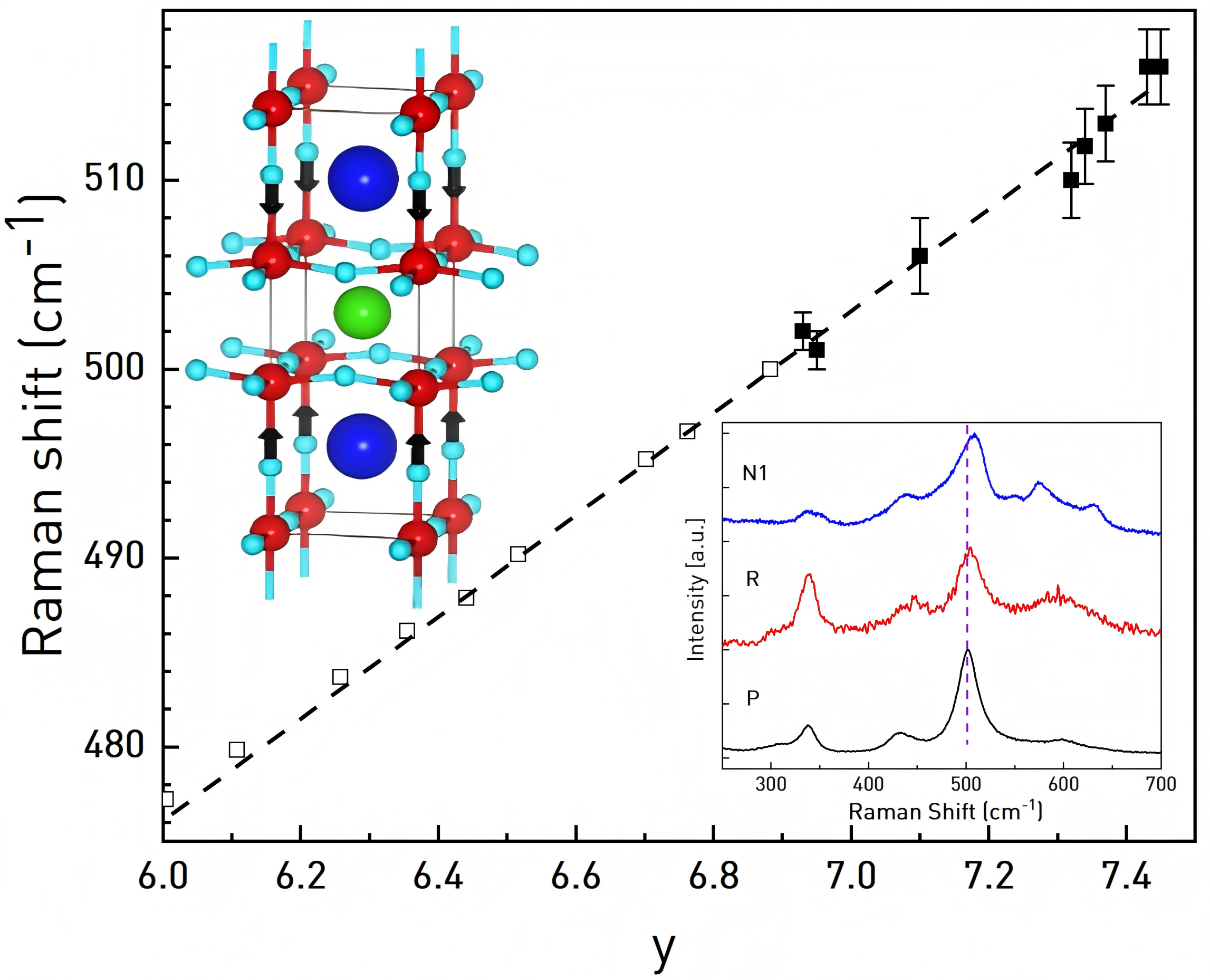} 
    \caption{Raman shift of the A$_{1g}$ Raman mode of \YBCO\ (see top-left figure) as a function of $y$. Open and full squares refer to the underdoped region, $y < y^{\ast}=6.9$ (data taken from Ref. \cite{tho88}), and to the overdoped ($y > y^{\ast}$) HPO samples studied in the present work, where $y$ was determined by structure refinement or estimated from the lattice parameters. Inset: room temperature Raman spectra of three representative samples: optimally doped \YBCOopt\ (P); sample P submitted to the HP treatment with no oxidizing agent (R); HPO sample N1. Note the main Raman modes of \YBCOopt\ at 154, 334, 438 and 502 cm$^{-1}$ (marked by a broken vertical line).}
    \label{fig:raman}
\end{figure}

As a third method to determine $y$, we used a remarkably linear dependence of the frequency of the A$_{1g}$ Raman mode on $y$ earlier established by Thomsen \textit{et al.} \cite{tho88} in the underdoped region, $y=6.0-6.9$ (see Figure \ref{fig:raman}). The mode consists of a stretching of the apical oxygen O1 along the $c$-axis, so its frequency is sensitive to the strength of the Cu1-O1 and Cu-O2 bonds. It was previously found \cite{cav90,jor90} that the former bond expands by 0.05 \AA\ while the latter bond shrinks by 0.20 \AA\ with increasing $y$ in the $y=6.0-6.9$ range. Since the second change dominates, the mode hardens at a rate of 25 cm$^{-1}$ / $y$, the frequency being 502 cm$^{-1}$ at optimal doping, $y^{\ast}=6.9$. Thus, for a typical resolution of 1 cm$^{-1}$, the accuracy of the estimate of $y$ is better than $\delta y=\pm 0.05$.

An independent determination of $y$ by means of the two previous methods, i.e. structure refinement and dependence of the lattice parameters on $y$, shows that the above linear relationship can be extrapolated in the overdoped region and thus to estimate $y$ up to $y=0.5$, as seen in Figure \ref{fig:raman}.

For an accurate measure of the A$_{1g}$ mode, care was taken to avoid laser-induced heating of the sample. This may occur in our HPO samples, owing to their limited thermal stability. TGA shows that excess oxygen starts to desorb at temperatures below 200 °C. For each sample, we have taken a series of spectra on several spots by varying beam power and acquisition time, i.e. 0.5 mW for 10 or 30 minutes and 0.5 mW or 5 mW for 2 minutes. We found that a 5 mW beam generates impurities visible as extra modes at 563 cm$^{-1}$, 590 cm$^{-1}$ and 633 cm$^{-1}$, attributed to Ba$_{2}$CuO$_{3}$, Y$_{2}$Cu$_{2}$O$_{5}$ and BaCuO$_{2}$ \cite{pop88,cha94}. Care was taken to limit exposure of the sample to air. After two hours of exposure, we observed an additional double peak near 340 cm$^{-1}$ and other minor features in the 100-200 cm$^{-1}$ range. Using the above precautions, a typical Raman spectrum of a HPO sample differs from that of optimally doped YBCO only in the frequency and relative intensities of the modes, as expected for a single-phase sample. This is seen in Figure \ref{fig:raman}, where we plot the spectra of three representative samples:

\vspace{4 pt}
\noindent
\textit{Sample P}. This is optimally doped YBCO used as precursor for the HPO treatment. Its spectrum, used as reference, is identical to previous spectra of optimally doped YBCO \cite{kir88,tho88}.

\vspace{4 pt}
\noindent
\textit{Sample R}. This is the sample obtained by treating the precursor P under the same HP conditions used to obtain the HPO samples, without oxidizing agent. Therefore, the oxygen concentration in this sample is expected to be unchanged, $y^{\ast}=6.9$. Indeed, its spectrum does not differ appreciably from that of the precursor P, except for a difference in the relative intensities of the modes, which we attribute to slight structural differences caused by quenching the sample under high pressure.



\vspace{4 pt}
\noindent
\textit{Sample N1 ($y=0.33$)}. Note a clear hardening of the A$_{1g}$ mode and an additional small peak at 632 cm$^{-1}$, previously attributed to a vibrational mode of Cu-O pairs in the CuO chains \cite{kir88}, consistent with the formation of fragments of CuO chain along the $a$-axis in presence of extra oxygens in the O5 site.   

In summary, the measurement of the A$_{1g}$ mode offers a straightforward and non-destructive method to estimate oxygen concentration in both under and over-doped YBCO.


\section{Structure changes induced by excess oxygen}
\begin{figure}[htbp]
    \centering
    \includegraphics[width=\columnwidth]{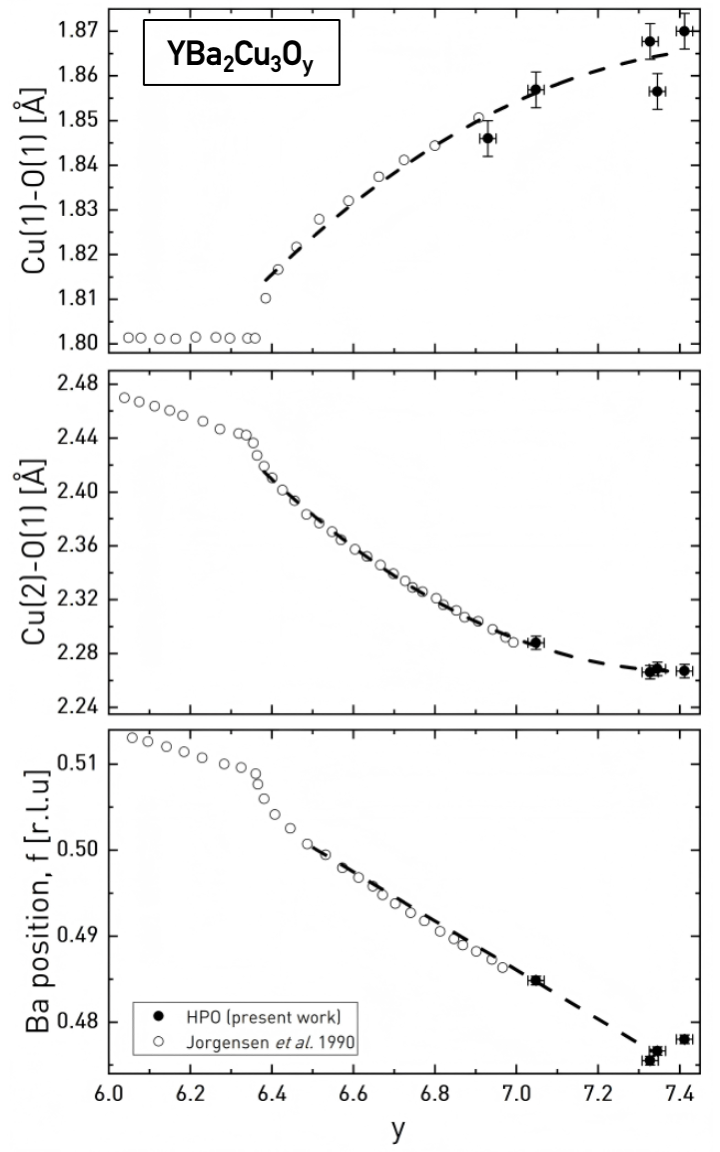} 
    \caption{Cu2- and Cu1-O1 bond lengths and Ba position, $f$, of \YBCO\ as a function of $y$. $f$ is the distance in reduced lattice units from the O4/O5 plane at $z=0$ to the average position of the Cu2-O2/O3 plane. Open and full symbols refer to underdoped samples (data from Jorgensen et al. \cite{jor90}) and to the overdoped N1-N4 samples studied in the present work. Error bars are taken from Table \ref{tab:HPO_ISIS}.} 
    \label{fig:parameters}
\end{figure}

In Tables \ref{tab:dist_ISIS} and \ref{tab:dist_SOLEIL} we report selected bond lengths taken from the refined structures of Tables \ref{tab:structure_ISIS} and \ref{tab:structure_SOLEIL} and plot the most relevant bond lengths as a function of $y$ in Figure \ref{fig:parameters}. Note that all atomic positions smoothly extrapolate the trend previously reported in the underdoped region \cite{cav90,jor90} up to the maximum value $y=0.5$ achieved in the present work.

We first focus on the bonds between the apical oxygen (O1 site) and the copper ions in chain (Cu1) and planar (Cu2) sites. As mentioned before, in the underdoped region, the Cu1-O1 distance increases monotonically with oxygen concentration, whereas the opposite occurs for the Cu2-O1 distance \cite{cav90,jor90}. The first trend is explained by an expansion of the $b$-axis parameter caused by the progressive occupation of the O4 site. The second trend reflects an increase in the valence of the planar Cu2 copper ion, as a result of the hole transfer from the CuO chain to the CuO$_2$ planes. In Figure \ref{fig:parameters}, we see that both trends continue smoothly in our HPO samples, i.e. the Cu1-O1 distance increases further with $y$, while the Cu2-O1 distance decreases, which gives a first indication of a (at least partial) transfer of extra holes to the CuO$_2$ planes.

The Ba position along the $c$-axis continues the trend in the underdoped region: it moves closer to the CuO chains as a result of the increase of coordination number from 9 at $y=7$ to 10 at $y=8$, which increases the electrostatic potential of the oxygen ions in the O4 and O5 sites. Concomitant to this change, the buckling of the Ba–O plane decreases with $y$.

The structural parameters in Tables \ref{tab:dist_ISIS} and \ref{tab:dist_SOLEIL} further indicate that the extra oxygen in the O5 site does not alter the CuO$_2$ planes, as expected considering that this site is located away from the planes. For example, the separation between CuO$_2$ planes increases in the underdoped region from 3.28 \AA\ in YBa$_2$Cu$_3$O$_6$ to 3.38 \AA\ in \YBCOopt\ and continues to increase slightly in our HPO samples until it levels off at 3.40 \AA. A further indication that the planes remain intact is that the Raman modes at 334 and 438 cm$^{-1}$, corresponding to the bending and stretching vibrations of the planes, respectively \cite{kuz89}, do not exhibit any significant changes.

In conclusion, HPO YBCO appears to be a model system to study the effects of overdoping disentangled from those of structural changes, contrary to the case of Ca-overdoping, as discussed in section \ref{Ca-YBCO}.

\section{Estimate of the overdoping level, $\Delta p$}
As said before, the shorter Cu2-O1 distance gives a first indication that a fraction of extra holes produced by the excess oxygen is transferred from the CuO chains to the CuO$_2$ planes. We now try to quantify the effective increase, $\Delta p$, of hole density by means of a bond valence sum (BVS) analysis. According to BVS theory \cite{bro09,bro85}, the valence state, $v$, of a given ion is the sum of bond valences of the ion:

\begin{equation}
v= \sum_i \exp \frac{R_0 - R_{i}}{B}
\end{equation}

where $R_{i}$ are the lengths of the bonds between the ion and the $i$-th ligand atom and the summation extends over all bonds. The parameter $B$ is an empirical constant, generally set at 0.37, and $R_0$ is the length of the bond for a given valence state and for a given coordination number, empirically determined from the structure of standard compounds.

Various authors applied the BVS method to YBCO. Here, we use the $R_0$ values proposed \textit{ad hoc} by Brown \cite{bro89} for underdoped YBCO, i.e. 1.679 \AA\ and 1.730 \AA\ for Cu$^{2+}$ and Cu$^{3+}$, respectively. Some authors \cite{moh05} suggested that the internal stress in the structure would require a correction of the value of $B$. However, this correction is expected to be small and, in any case, we are interested in the variation of $p$ rather than its absolute value, so we keep the usual value, $B=0.37$. For both Cu1 and Cu2 sites, we calculate $v$ using a weighted average of the above $R_0$ values for Cu$^{2+}$ and Cu$^{3+}$ ions and checking self-consistency. We apply the method using the refined structural parameters of the N1-N4 samples reported in Table \ref{tab:structure_ISIS}.

In Figure \ref{fig:BVS_V}a, we compare our result with the results reported by Brown \cite{bro89}. Consistent with the previous finding that the structural changes in our HPO samples smoothly follow the trend in the underdoped region, Figure \ref{fig:BVS_V}a shows that the same occurs for the valence of the Cu1 and Cu2 sites. Namely, the valence of the Cu1 site continues to increase linearly with $y$ up to 3+, while the valence of the Cu2 site remains constant at the value 2.2+ of optimally doped samples. At first sight, this may indicate that the extra holes produced by the excess oxygens $y$ are retained in the charge reservoir layer of the CuO chains. In fact, holes in the CuO$_2$ planes are known to be located mainly in the ligand at the O2 and O3 sites, so Tallon \cite{tal90} proposed that a realistic measure of $p$ is the \textit{total} bond valence sum, $V^{-}$:




\begin{figure}[hptb!]
    \centering
    \includegraphics[width=\columnwidth]{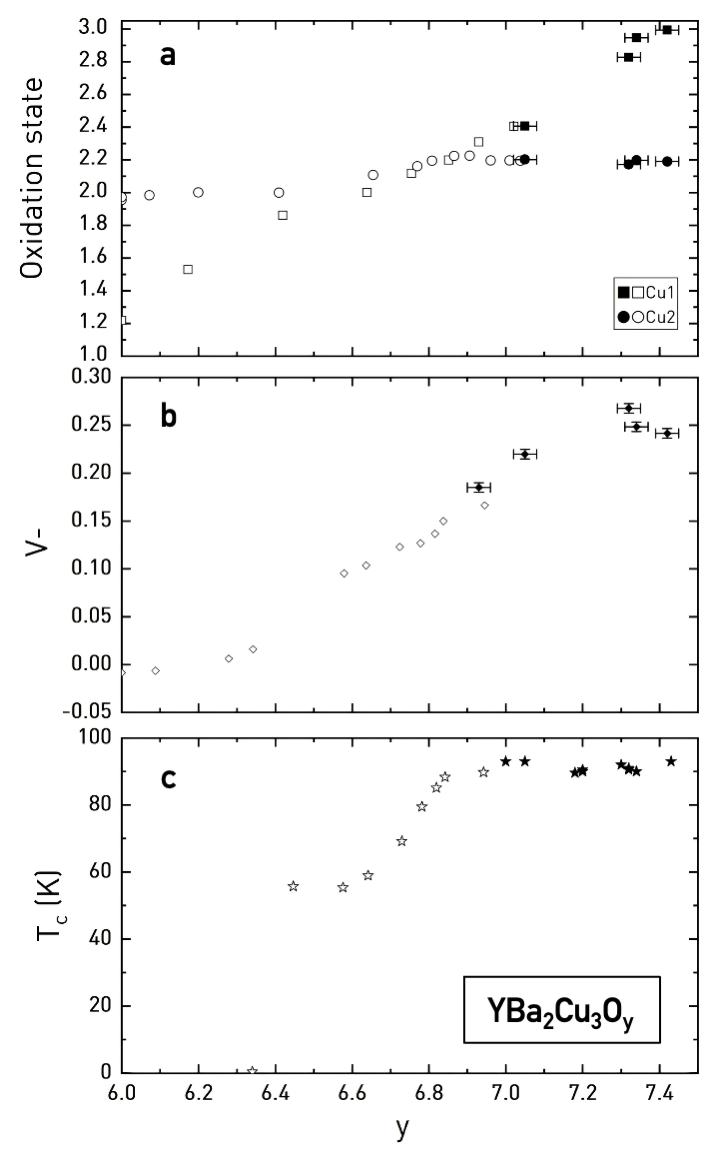} 
    \caption{a): Oxidation state of the Cu1 and Cu2 sites, obtained from the BVS analysis described in the text, vs. oxygen content, $y$. Open symbols refer to under- and optimally doped YBCO (data from Brown \cite{bro89}); full symbols are obtained from the refined structures of samples HPO N1-N4 in Table \ref{tab:structure_ISIS}. b): $y$-dependence of the total bond valence sum, $V^{-}$ (see Eq. \ref{eq:V-}), proposed by Tallon \cite{tal90} as a measure of hole density, $p$, in the CuO$_2$ plane. Open symbols refer to under- and optimally doped YBCO from Tallon \cite{tal90}; full symbols refer to samples P and HPO N1-N4. c): $y$-dependence of the onset critical temperature, $T_c$. Open symbols are taken from Cava et al. \cite{cav90}; full symbols refer to our P and HPO samples.} 
    \label{fig:BVS_V}
\end{figure}

\begin{equation}
V^{-} = 2 + V_{\rm Cu2} - V_{\rm O2} - V_{\rm O3}
\label{eq:V-}
\end{equation}

which represents the sum of the excess positive charge on the Cu2 ion, $V_{\rm Cu2} - 2$, and on the O2 and O3 oxygen ions, $2-V_{\rm O}$. As seen in Figure \ref{fig:BVS_V}b, our HPO samples display a sizeable increase of $V^{-}$ with $y$ from the optimal value 0.17 up to 0.27. By taking $V^{-}$ as a measure of $p$, we thus conclude that hole doping reaches the border of the overdoped side of the dome, where superconductivity is expected to vanish.

\section{Superconducting properties}
\begin{figure}[htbp]
\centering
\includegraphics[width=\columnwidth]{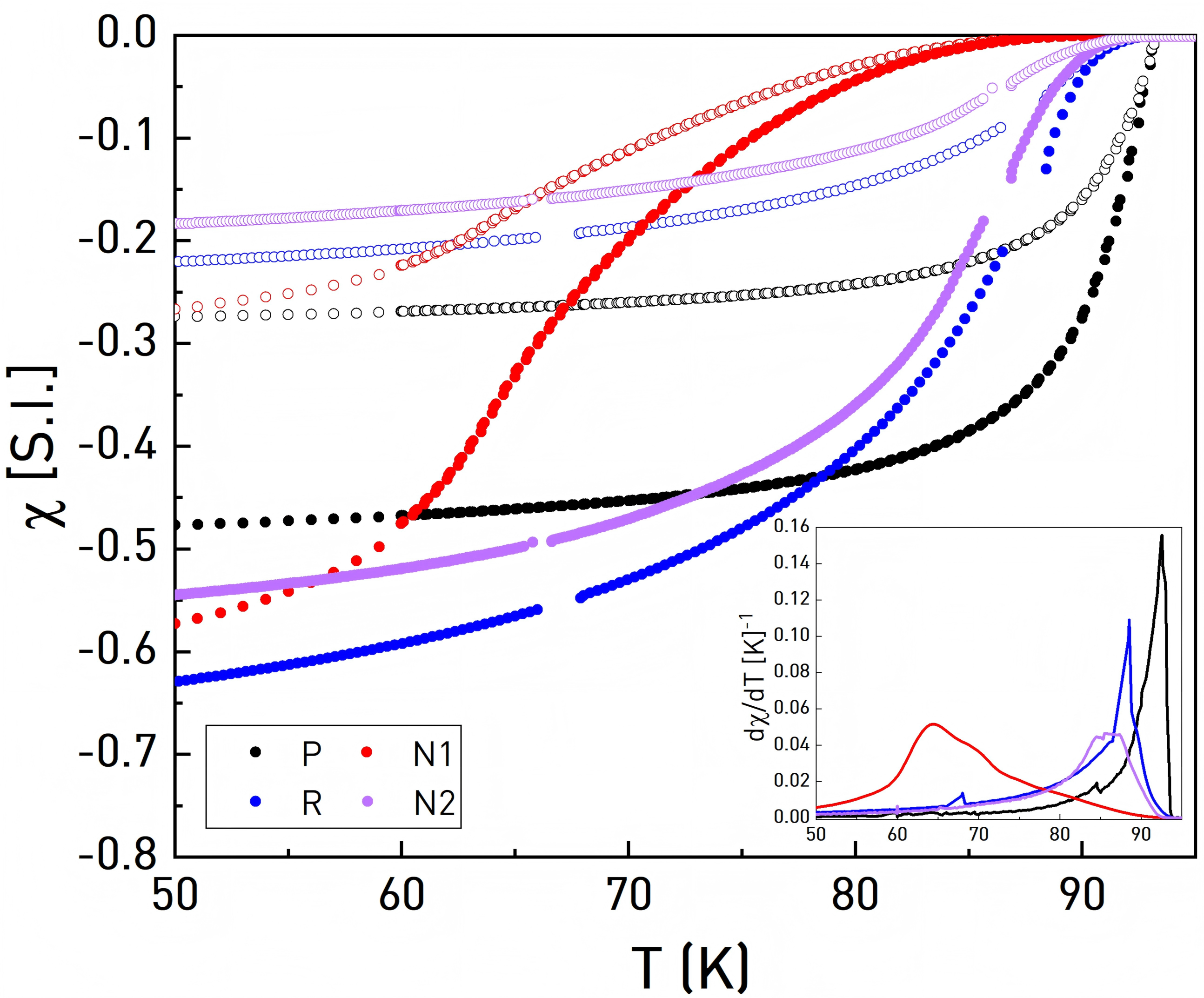} 
\caption{Magnetic susceptibility of sample P (optimally doped \YBCOopt\ used as precursor for the HPO synthesis), sample R (sample P after the HP treatment without oxidizing agent) and the HPO samples N1 and N2 prepared using KClO$_3$ and AgO as oxidizing agent, respectively (see Table \ref{tab:HPO_ISIS}). Full and open symbols refer to ZFC and FC curves. Inset: derivative curves.}
\label{fig:chi}
\end{figure}

Contrary to the above expectation and in agreement with earlier results by Okai et al. \cite{oka02}, our susceptibility results in Figures \ref{fig:BVS_V}c and \ref{fig:chi} show that, in the whole $y$-range studied, the onset critical temperature, $T_c$, does not change significantly with respect to the $\sim 92$ K value of optimally doped samples. The superconducting fraction, defined as the low-field susceptibility in the FC curve, does not change either, $f \sim 0.2-0.3$, which supports the observation of single- or nearly single-phase samples and indicates that superconducting is a bulk property. Different from us, Okai et al. \cite{oka90,oka96,oka99} reported a large reduction of $f$ with increasing $y$ down to $f \sim 0.1$ for oxygen compositions $y \sim 0.5$ similar to ours. This difference is attributed to the formation of impurity phases at the high synthesis temperatures up to 1150 °C employed by Okai et al., while the lower temperatures used here prevent this issue.

The susceptibility curves of the HPO samples differ from those of optimally doped YBCO in a broadened transition (see Figure \ref{fig:chi}). The broadening is more evident in the HPO samples prepared using KClO$_{3}$, which suggests that this effect is related to the microstructure. Indeed, the oxidizing power of KClO$_{3}$ is larger than that of AgO, so the alteration of the microstructure of the superconducting grains is expected to more pronounced. In the absence of impurity phases and of multiple transitions, which is better seen in the derivative curves, $d\chi/dT$, we attribute the broadening to a large density of defects caused by the high-pressure diffusion of oxygen in the grains. As a result, by approaching the transition, the magnetic penetration depth, $\lambda(T)$, becomes comparable or larger than the size of defect-free grains. The screening currents then occupy a large fraction of the grain and a full diamagnetic response occurs only far from the transition, where $\lambda(T)$ is minimum.  

Finally, it is likely that oxygen disorder causes a further broadening. Indeed, a sizable broadening is found in underdoped YBCO as well \cite{cav90}, where configurational disorder of the O4 and O5 sites is large, especially for $y=6.5$, similar to the present case, $y=7.5$. It is likely that the quenching under high pressure increases oxygen disorder in our HPO samples, as suggested by the fact that a sizable - although less pronounced - broadening is seen in the optimally doped sample R prepared under high pressure without oxidizing agent.

\section{Discussion}

\subsection{Oxygen vs. Ca overdoping}
\label{Ca-YBCO}

The present structure analysis enables us to explain the opposite effects on $T_c$ produced by oxygen vs. Ca overdoping. We recall that, in \YCaBCO, the partial Ca-substitution quickly lowers $T_c$ down to 56 K for $x=0.2$, concomitant to an overdoping level, $p \approx 0.2-0.3$ hole/Cu \cite{bot96}, estimated using the same BVS analysis used above. In fact, contrary to the case of oxygen overdoping, Ca substitution alters significantly the crystal structure of the CuO$_2$ planes \cite{bot96} and thus the electronic properties, owing to the larger size and smaller charge of the Ca$^{2+}$ ion as compared to Y$^{3+}$. The main change is an increase of the Cu2–O1 bond length with $x$ \cite{bot96}, while this length decreases with oxygen concentration (see Figure \ref{fig:parameters}). This structural change alone may explain the decrease of $T_c$, considering that a long Cu2–O1 bond hinders the transfer of holes to the CuO$_2$ layers \cite{cav90,jor90}. Further changes induced by Ca substitution are a reduction of the separation between the two CuO$_2$ planes and of the buckling of the planes \cite{bot96}.   
In conclusion, it is not straightforward to establish whether the $T_c$ reduction in \YCaBCO\ is caused by hole overdoping, as expected according to the dome scenario, or by the structural changes in the CuO$_2$ layers. So, we argue that \YBCO\ is a more suitable system to study the effects of hole overdoping disentangled from those of structural changes. 



\subsection{Effects of oxygen disorder}
Since in our HPO samples the excess oxygen partially occupies the O5 site, the question arises whether the occupancy is random or ordered. The effectiveness of hole transfer from the CuO chains to the CuO$_2$ planes is expected to be different in these two cases. According to early reports \cite{cav88,cav90,bur92}, sufficiently long segments of CuO chains are needed for an effective transfer. This scenario is supported by comparing the $T_c$ of optimally doped \YBCOopt\ with its counterpart \YSCO\ (YSCO), where Sr substitutes Ba \cite{gil00}. It turns out that this substitution favors a disordered occupancy of the oxygens in the O4 and O5 sites, thus stabilizing a tetragonal P4/$mmm$ symmetry. Consistent with the observation that the maximum $T_c$ is significantly lower in YSCO as compared to YBCO, a first-principles theoretical study \cite{gau19} suggests that the disorder hinders charge transfer.

By applying the above considerations to the present HPO samples, we envisage that the extra oxygens in the O5 site either form segments of CuO chains along the $a$-axis, and thus an additional electronic band, or randomly occupy the O5 site. In the former case, a full CuO chain would be formed for $y\sim0.5$, thus leading to a structural modulation, but our diffraction data do not give any indication of long-range order. In conclusion, the ordering, if any, may only occur in regions smaller than the diffraction coherence length, $\sim 100$ nm. The comparatively large thermal factors, $B_{\rm iso}$, of the O4 and O5 sites (see Table \ref{tab:structure_ISIS}) rather suggest a pronounced disorder, which may, at least partially, explain the broadening of the transition, similar to the case of underdoped \YBCO\ for $y\sim 6.5$, where the O4 and O5 sites are also disordered. Further studies of the local structure, such as pair distribution function (PDF) analysis or NMR and NQR spectroscopy, may elucidate this point.

\subsection{Two-band scenario}
Our results confirm earlier evidence of an extended superconducting region in the overdoped region of some cuprates, which contrasts the dome scenario. In the absence of a theory of superconductivity for cuprates, such an apparently contradicting behavior of cuprates indicates that the usual single-band approximation, where $p$ represents the number of $b_1$-symmetry Zhang-Rice singlets \cite{zha88} formed by a hole in the Cu 3$d_{x^2-y^2}$ orbital and a hole in the O 2$p_x$ and $p_y$ orbitals, is no longer appropriate for the overdoped regime. Indeed, the approximation is valid in the lightly doped regime, while X-ray Absorption Spectroscopy revealed that the singlets become unstable at high hole densities \cite{pee09}.


The need for a more complex model was earlier noted by Di Castro, Feiner and Grilli \cite{cas91} who proposed that a second hole state with $a_1$ symmetry, formed by the Cu 3$d_{3z^2-r^2}$ orbital, must be taken into account to explain the large $T_c$ variations observed in cuprates with similar doping level, $p$. Remarkably, the occupancy of this orbital is significant in all systems that exhibit superconductivity in the heavily overdoped region beyond the dome, owing to an unusually short distance between the apical oxygen and the planar Cu ion \cite{chm10,mar13b,gau16,li19,cam25}. It was then suggested \cite{mai19,sca19} that, in these systems, the above orbital forms a second band that contributes to superconductivity.

In the light of the above considerations, we put forward the hypothesis that, in overdoped \YBCO\ as well as in other cuprate systems that display an extended superconducting region at high doping, the overdoping provided by the extra oxygens alters the relative occupancy of hole states with $a_1$ or $b_1$ symmetries. Within this scenario, the $T_c$ stability is explained by the fact that the extra holes mainly occupy the $a_1$ states. On the contrary, in the overdoped cuprates that follow the dome, the extra holes mainly occupy the $b_1$-symmetry Zhang-Rice singlet states that become unstable, which would explain the suppression of superconductivity in these systems.


\section{Conclusion}
The present work on high-pressure oxygenated \YBCO\ shows that a large excess of oxygen concentration, $y$ up to 7.4, with respect to optimal doping, $y^{\ast}=6.9$, is effective in achieving overdoping of the CuO$_2$ planes up to $p=0.27$ hole/Cu without reducing $T_c$, which contrasts the dome scenario. To explain this discrepancy, our structure analysis suggests that the extra holes increases the occupancy of hole states with $a_1$ symmetry formed by the $d_{3z^2-r^2}$ orbital, without affecting the occupancy of the Zhang-Rice singlet states with $b_1$ symmetry formed by the $d_{x^2-y^2}$ orbital. This points at the inadequacy of the single-band Zhang-Rice singlet approximation to describe the overdoped regime of cuprates. In order to test this two-band scenario, we propose extending the present study to HPO single crystals, which would enable to probe the electronic and transport properties of YBCO as well as other cuprates in the strongly overdoped regime.




\begin{acknowledgments}
The authors thank J.P. Attfield, J.K. Byland, C. Decorse, L. Forr\'o, M. Grilli, B. Keimer, G. Kim, S.A. Kivelson, T. Loew, D. Pavuna, J. R\"ohler, A. Sacuto, G.A. Sawatzky, H. Takagi, J.L. Tallon, Y. Uemura and I. Vishik for useful discussions, D. Deldicque, O. Beyssac and G. Baldinozzi for assistance in the Raman measurements, H. Moutaabbid for assistance in sample preparation and gratefully acknowledge the ANR for financial support under contract ANR-22-CE30-0010, SUPERSTRONG. Data collection at the ISIS Neutron and Muon Source was supported by beamtime allocation RB2420252 from the Science and Technology Facilities Council. Neutron Diffraction Data are available from https://doi.org/10.5286/ISIS.E.RB2420252-1.
\end{acknowledgments}

\appendix

\section{Sample synthesis conditions}
\label{app:synthesis}
In Table \ref{tab:HPO_ISIS} we give a summary of the synthesis conditions of the N1-N4 samples studied by neutron diffraction (see Appendix \ref{app:refined}). The oxidizing agent was either mixed with the \YBCOopt\ precursor powder (samples N1 and N3) or separated by a perforated Au foil (samples N2 and N4). All samples were prepared with a concentration of nominal oxidizer corresponding to a YBa$_2$Cu$_3$O$_{8.25}$ composition, i.e. with an excess oxygen as compared to the maximum concentration $y=8$ that can be hosted by the YBCO structure. In Table \ref{tab:HPO_SOLEIL}, we give the same summary for the S1-S4 samples studied by synchrotron X-ray diffraction (see Appendix \ref{app:refined}). As before, the oxidizing agent was either mixed with the \YBCOopt\ precursor powder (samples S2, S3 and S4) or separated by a perforated Au foil (sample S1).\\

\begin{table*}[htbp]
\centering
\setlength{\tabcolsep}{6pt}
\renewcommand{\arraystretch}{1.3}
  \caption{Summary of the synthesis conditions of the N1-N4 samples studied by neutron diffraction.}
  \label{tab:HPO_ISIS}
\begin{tabular}{lllll}
\hline \hline
Sample name & N1 & N2 & N3 & N4 \\
\hline \hline
Oxidizing agent & KClO$_3$ & AgO & KClO$_3$ & KClO$_3$ \\
\hline
Oxidizer ratio & 1:0.412 & 1:0.412 & 1:0.412 & 1:0.412\\
\hline
Synthesis pressure & 6 GPa & 6 GPa & 6 GPa & 4 GPa \\
\hline
Synthesis temperature & 475°C &  475°C &  550 °C & 475 °C \\
\hline
Synthesis time & 3.5h & 3.5h & 3.5h & 2h \\
\hline
Cooling & Quench & Quench & Quench & 350 °C/min \\
\hline
Impurity concentration & $<1\%$ & $<2\%$ & $<2\%$ & $<1\%$ \\
\hline \hline
\end{tabular}
\end{table*}

\begin{table*}[htbp]
\centering
\setlength{\tabcolsep}{8pt}
\renewcommand{\arraystretch}{1.3}
\caption{The same as in Table \ref{tab:HPO_ISIS} for the S1-S4 samples studied by synchrotron X-ray diffraction.}
\label{tab:HPO_SOLEIL}

\begin{tabular}{lllll}
\hline \hline
Sample name & S1 & S2 & S3 & S4\\
\hline \hline
Oxidizing agent & AgO & KClO$_3$ & KClO$_3$ & KClO$_3$\\
\hline
Oxidizer ratio & 1:0.412 & 1:0.412 & 1:0.5 & 1:0.333\\
\hline
Synthesis pressure & 6 GPa & 6 GPa & 6 GPa & 6 GPa\\
\hline
Synthesis temperature &  475°C &  475°C & 475°C & 350 °C\\
\hline
Synthesis time & 120h & 120h & 3.5h & 80h\\
\hline
Cooling & Quench & Quench & Quench & Quench\\
\hline
Impurity concentration & $<5\%$ & $<1\%$ & $<5\%$ & $<5\%$ \\
\hline \hline
\end{tabular}
\end{table*}

\section{Refined structures and selected bond lengths}
\label{app:refined}

In Table \ref{tab:structure_ISIS} we give the refined crystal structure of the N1-N4 samples measured in the POLARIS diffractometer of the ISIS Neutron and Muon Source. For comparison, we give the refined structure of optimally doped \YBCOopt\ \cite{jor90}. In Table \ref{tab:structure_SOLEIL} we give the refined crystal structure of the S1-S4 samples measured at the CRISTAL beamline of the SOLEIL synchrotron facility. The data were collected at constant wavelength, $\lambda =0.51363$ \AA. In Tables \ref{tab:dist_ISIS} and \ref{tab:dist_SOLEIL}, we give the values of selected bond lengths extracted from the refined structures of Tables \ref{tab:structure_ISIS} and \ref{tab:structure_SOLEIL}, respectively. 

\renewcommand{\arraystretch}{1.4}
\setlength{\tabcolsep}{8pt}


\begin{table*}[htbp]
\centering
\caption{Refined structure in the $Pmmm$ symmetry of the N1-N4 samples measured at room temperature in the POLARIS diffractometer. Site labels are as in Figure \ref{fig:structure} and in \cite{cap87}. We refined the site occupancy factor (SOF) of the O5 site by keeping fixed to 1 the SOF of the O4 site. Numbers in parentheses indicate statistical uncertainty. For comparison, we give the refined structure of optimally doped \YBCOopt\ \cite{jor90}.}

\label{tab:structure_ISIS}

\begin{tabular}{llllll}
\hline \hline
& YBa$_2$Cu$_3$O$_{6.9}$ \cite{jor90} & Sample N1 & Sample N2 & Sample N3 & Sample N4\\
\hline \hline
$a$(\AA) & 3.8227(1) & 3.8458(2) &  3.8271(2) & 3.8469(1) &  3.8310(9)\\
\hline
$b$(\AA) & 3.8872(2) & 3.8858(2) &  3.8837(2) & 3.8797(3) &  3.8837(1)\\
\hline
$c$(\AA) & 11.6802(2) & 11.6541(5) &  11.6750(5) & 11.6441(7) &  11.6470(9)\\
\hline
$V$(\AA)${^3}$ & 173.560(4) & 174.159(4) & 173.529(6) & 173.786(1) & 173.289(4) \\
\hline \hline
\multicolumn{6}{l}{\textbf{Y} $1h$ (\textonehalf,\textonehalf,\textonehalf)} \\[-1ex]
$B_{iso}$(\AA)${^2}$ & 0.28(3) & 0.32(4) & 0.30(1) & 0.47(2) & 0.31(1)\\
\hline
\multicolumn{6}{l}{\textbf{Ba} $2t$(\textonehalf,\textonehalf,$z$)} \\[-1ex]
$z$ & 0.1843(2) & 0.1798(2) & 0.1822(3) & 0.1814(3) & 0.1802(1)\\[-1ex]
$B_{iso}$(\AA)${^2}$ & 0.44(3) & 0.50(6) & 0.40(2) & 0.50(5) & 0.48(2)\\
\hline
\multicolumn{6}{l}{\textbf{Cu1} $1a$ (0,0,0)}\\[-1ex]
$B_{iso}$(\AA)${^2}$ & 0.41(3) & 0.57(3) & 0.32(4) & 0.63(2) & 0.34(8)  \\
\hline
\multicolumn{6}{l}{\textbf{Cu2} $2t$ (0,0,$z$)}\\[-1ex]
$z$ & 0.3556(1) & 0.3547(1) & 0.3550(1) & 0.3553(2) & 0.3542(1) \\[-1ex]
$B_{iso}$(\AA)${^2}$ & 0.20(2) & 0.19(3) & 0.24(1 )& 0.08(1) & 0.24(8) \\
\hline
\multicolumn{6}{l}{\textbf{O1} $2t$ (0,0,$z$)} \\[-1ex]
$z$ & 0.1590(2) & 0.1603(1) & 0.1591(2) & 0.1606(3) & 0.1594(1) \\[-1ex]
$B_{iso}$(\AA)${^2}$ & 0.68(5) & 0.70(2) & 0.54(5) & 0.75(1) &  0.59(1) \\
\hline
\multicolumn{6}{l}{\textbf{O2} $2s$ (\textonehalf,0,$z$)} \\[-1ex]
$z$ & 0.3779(2) & 0.3788(3) & 0.3758(3) & 0.3795(2) & 0.3778(2) \\[-1ex]
$B_{iso}$(\AA)${^2}$& 0.51(4) & 0.42(1) & 0.44(2) & 0.29(2) & 0.37(3) \\
\hline
\multicolumn{6}{l}{\textbf{O3} $2r$ (0,\textonehalf,$z$)} \\[-1ex]
$z$ & 0.3790(2) & 0.3772(3) & 0.3799(1) & 0.3762(1) & 0.3781(2) \\[-1ex]
$B_{iso}$(\AA)${^2}$& 0.35(3) & 0.48(2) & 0.40(3) & 0.43(4) & 0.42(1)\\
\hline
\multicolumn{6}{l}{\textbf{O4} $1e$ (0,\textonehalf,0)} \\[-1ex]
SOF & 0.90(1) & 1  & 1 & 1  & 1 \\[-1ex]
$B_{iso}$(\AA)${^2}$& 1.1(1) & 4.49(3) & 1.42(9) & 7.18(5) & 4.99(1) \\
\hline
\multicolumn{6}{l}{\textbf{O5} $1b$ (\textonehalf,0,0)} \\[-1ex]
SOF & 0.03(1) & 0.327(5) & 0.048(1) & 0.415(4)  & 0.345(7) \\[-1ex]
$B_{iso}$(\AA)${^2}$& N.A. & 0.08 & 0.08 & 0.08 & 0.08 \\
\hline \hline
\multicolumn{6}{l}{Reliability factors} \\[-1ex]
$wR_p$  & 0.0596 & 0.0352 & 0.0380 & 0.0412 & 0.0218\\[-1ex]
$R_p$   & 0.0333 & 0.0334 & 0.0364 & 0.0378 & 0.0205 \\
\hline
No of free param. & N.A. & 58 & 51 & 56 & 51 \\
\hline \hline

\end{tabular}
\end{table*}


\renewcommand{\arraystretch}{1.4}
\setlength{\tabcolsep}{8pt}
\begin{table*}[htbp]
\centering
\caption{Refined structure in the $Pmmm$ symmetry of the S1-S4 samples measured at room temperature at the CRISTAL diffractometer of the SOLEIL synchrotron facility. The data were collected at constant wavelength, $\lambda =0.51363$ \AA. Site labels are as in Figure \ref{fig:structure} and in \cite{cap87}. We estimated the site occupancy factor (SOF) of the O5 site from the dependence of the lattice parameters on $y$ or from the shift of the Raman A$_{1g}$ mode (see text and Figures \ref{fig:abc_vs_y} and \ref{fig:raman}).}

\label{tab:structure_SOLEIL}
\begin{tabular}{lllll}
\hline \hline
         & S1        & S2        & S3        & S4 \\
\hline \hline
$a$(\AA) & 3.8387(1) & 3.8502(2) & 3.8328(5) & 3.8229(5)\\
\hline
$b$(\AA) & 3.8792(2) & 3.8838(2) & 3.8865(7) & 3.8866(6)\\
\hline
$c$(\AA) & 11.6668(5) & 11.6574(6) & 11.6534(2) & 11.6783(2)\\
\hline
$V$(\AA)${^3}$ & 173.731(10) & 174.319(14) & 173.592(5) & 173.522(4) \\
\hline \hline
\multicolumn{5}{l}{\textbf{Y} $1h$ (\textonehalf,\textonehalf,\textonehalf)} \\[-1ex]
$B_{iso}$(\AA)${^2}$ & 0.079 & 0.079 & 0.079 & 0.079\\
\hline
\multicolumn{5}{l}{\textbf{Ba} $2t$ (\textonehalf,\textonehalf,$z$)} \\[-1ex]
$z$ & 0.1823(1) & 0.1827(1) & 0.1804(1) & 0.1834(1)\\[-1ex]
$B_{iso}$(\AA)${^2}$ & 0.25(3) & 0.43(5) & 0.28(5) & 0.14(1) \\
\hline
\multicolumn{5}{l}{\textbf{Cu1} $1a$ (0,0,0)}\\[-1ex]
$B_{iso}$(\AA)${^2}$ & 0.39(6) & 0.59(3) & 0.37(9) & 0.26(1) \\
\hline
\multicolumn{5}{l}{\textbf{Cu2} $2t$ (0,0,$z$)}\\[-1ex]
$z$ & 0.3548(2) & 0.3534(1) & 0.3538(10) & 0.3556(8) \\[-1ex]
$B_{iso}$(\AA)${^2}$ & 0.02(2) & 0.09(1) & 0.31(5) &  0.09(1)\\
\hline
\multicolumn{5}{l}{\textbf{O1} $2t$ (0,0,$z$)} \\[-1ex]
$z$ & 0.1616(7) & 0.1606(1) & 0.1592(6) & 0.1600(4) \\[-1ex]
$B_{iso}$(\AA)${^2}$ & 0.21(8) &  0.47(5) & 0.82(9) &  0.28(4)  \\
\hline
\multicolumn{5}{l}{\textbf{O2} $2s$ (\textonehalf,0,$z$)} \\[-1ex]
$z$ & 0.3851(9) & 0.3855(2) & 0.3833(5) & 0.3801(4) \\[-1ex]
$B_{iso}$(\AA)${^2}$ & 0.21(8) & 0.47(5) & 0.82(9) & 0.28(4) \\
\hline
\multicolumn{5}{l}{\textbf{O3} $2r$ (0,\textonehalf,$z$)} \\[-1ex]
$z$ & 0.3764(10) & 0.3769(2) & 0.3746(6) & 0.3782(5) \\[-1ex]
$B_{iso}$(\AA)${^2}$ & 0.21(8) & 0.47(5) & 0.82(9) & 0.28(4) \\
\hline
\multicolumn{5}{l}{\textbf{O4} $1e$ (0,\textonehalf,0)} \\[-1ex]
$B_{iso}$(\AA)${^2}$ & 0.21(8) & 0.47(5) & 0.82(9) & 0.28(4)  \\
\hline
\multicolumn{5}{l}{\textbf{O5} $1b$ (\textonehalf,0,0)} \\[-1ex]
SOF & 0.1 & 0.3 & 0.35 & 0.05 \\[-1ex]
$B_{iso}$(\AA)${^2}$ & 0.21(8) & 0.47(5) & 0.82(9) & 0.28(4)  \\
\hline \hline
\multicolumn{5}{l}{Reliability factors} \\[-1ex]
$wR_p$ & 0.0617 & 0.0515 & 0.0566 & 0.0378 \\[-1ex]
$R_p$  & 0.0673 & 0.0699 & 0.0621 & 0.0437 \\
\hline
No. of free param. & 23 & 29 & 32 & 29 \\
\hline \hline
\end{tabular}
\end{table*}


\setlength{\tabcolsep}{12pt}

\begin{table*}[htbp]
\centering
\caption{Selected bond distances in \AA\ and bond angles in $^\circ$ corresponding to the refined structure of Samples N1-N4 in Table \ref{tab:structure_ISIS}. Numbers in parentheses indicate statistical uncertainty.}

\label{tab:dist_ISIS}
\begin{tabular}{lccccc}
\hline \hline
 & \YBCOopt & Sample N1 & Sample N2 & Sample N3 &  Sample N4 \\
\hline \hline
\multicolumn{4}{l}{\textit{Bond lengths}} \\
Cu1-O1 & 1.846(1) & 1.868(2) & 1.857(2) & 1.870(3) & 1.857(1) \\
Cu2-O1 & 2.298(7) & 2.266(2) & 2.288(3) & 2.267(3) & 2.269(1) \\
Cu2-O2 & 1.929(1) & 1.943(5) & 1.929(5) & 1.944(1) & 1.935(1) \\
Cu2-O3 & 1.961(1) & 1.961(1) & 1.963(1) & 1.955(1) & 1.962(1) \\
Ba1-O1 & 2.741(1) & 2.743(2) & 2.740(2) & 2.743(1) & 2.738(1) \\
Ba1-O4 & 2.876(4) & 2.844(2) & 2.861(2) & 2.857(3) & 2.738(1) \\
Cu2-Cu2 & 3.388(1) & 3.386(3) & 3.385(3) & 3.371(4) & 3.397(2)\\
\hline \hline
\multicolumn{4}{l}{\textit{Bond angles}} \\
Cu2-O2-Cu2 & 164.0(1) & 163.4(2) & 165.5(2) & 163.3(4) & 163.7(1) \\
Cu2-O3-Cu2 & 164.4(1) & 164.6(2) & 163.0(2) & 165.7(3) & 163.7(1) \\
\multicolumn{4}{l}{\textit{Polyhedral Volume}} \\
Cu2 site & 6.349(1) & 6.344(1) & 6.387(1) & 6.341(1) & 6.317(1) \\
\hline \hline 

\end{tabular}
\end{table*}


\setlength{\tabcolsep}{12pt}

\begin{table*}[htbp]
\centering
\caption{Selected bond distances in \AA\ and bond angles in $^\circ$ corresponding to the refined structure of samples S1-S4 taken from Table \ref{tab:structure_SOLEIL}. Numbers in parentheses indicate statistical uncertainty.}
\label{tab:dist_SOLEIL}
\begin{tabular}{lcccc}
\hline \hline
 & S1 & S2 & S3 & S4 \\
\hline \hline
\multicolumn{3}{l}{\textit{Bond lengths}} \\
Cu1-O1 & 1.885(9) & 1.872(1) & 1.855(7) & 1.869(5) \\
Cu2-O1 & 2.253(9) & 2.247(1) & 2.268(8) & 2.284(5) \\
Cu2-O2 & 1.952(2) & 1.961(3) & 1.9470(11) & 1.933(1) \\
Cu2-O3 & 1.956(16) & 1.961(2) & 1.9583(9) & 1.961(1) \\
Ba1-O1 & 2.739(1) & 2.741(1) & 2.7428(6) & 2.740(1) \\
Ba1-O4 & 2.865(1) & 2.845(1) & 2.8451(6) & 2.871(2) \\
Cu2-Cu2 & 3.389(4) & 3.408(3) & 3.422(3) & 3.374(2)\\
\hline \hline
\multicolumn{3}{l}{\textit{Bond angles}} \\
Cu2-O2-Cu2 & 159.1(7) & 158.0(9) & 159.7(4) & 162.9(3) \\
Cu2-O3-Cu2 & 165.2(7) & 163.9(9) & 165.8(5) & 164.5(4) \\
\hline \hline 
\multicolumn{3}{l}{\textit{Polyhedral Volume}} \\
Cu2 site & 6.472(1) & 6.534(2) & 6.484(2) & 6.365(5) \\
\hline \hline 

\end{tabular}

\end{table*}


\bibliography{biblio}

\end{document}